# Beyond Low Earth Orbit: Biological Research, Artificial Intelligence, and Self-Driving Labs

Lauren M. Sanders[1] , Jason H. Yang[2], Ryan T. Scott[3], Amina Ann Qutub[4], Hector Garcia Martin[5,6,7], Daniel C. Berrios[3], Jaden J.A. Hastings[8], Jon Rask[9], Graham Mackintosh[10], Adrienne L. Hoarfrost[11], Stuart Chalk[12], John Kalantari[13], Kia Khezeli[13], Erik L. Antonsen[14], Joel Babdor[15], Richard Barker[16], Sergio E. Baranzini[17], Afshin Beheshti[3], Guillermo M. Delgado-Aparicio[18], Benjamin S. Glicksberg[19], Casey S. Greene[20], Melissa Haendel[21], Arif A. Hamid[22], Philip Heller[23], Daniel Jamieson[24], Katelyn J. Jarvis[25], Svetlana V. Komarova[26], Matthieu Komorowski[27], Prachi Kothiyal[28], Ashish Mahabal[29], Uri Manor[30], Christopher E. Mason[8], Mona Matar[31], George I. Mias[32], Jack Miller[3], Jerry G. Myers, Jr.[31], Charlotte Nelson[17], Jonathan Oribello[1], Seung-min Park[33], Patricia Parsons-Wingerter[34], R. K. Prabhu[35], Robert J. Reynolds[36], Amanda Saravia-Butler[37], Suchi Saria[38,39], Aenor Sawyer[24], Nitin Kumar Singh[40], Frank Soboczenski[41], Michael Snyder[42], Karthik Soman[17], Corey A. Theriot[43,44], David Van Valen[45], Kasthuri Venkateswaran[40], Liz Warren[46], Liz Worthey[47], Marinka Zitnik[48], Sylvain V. Costes[49+]

[1] Blue Marble Space Institute of Science, Space Biosciences Division, NASA Ames Research Center, Moffett Field, CA 94035, USA.
[2] Center for Emerging and Re-Emerging Pathogens, Department of Microbiology, Biochemistry and Molecular Genetics, Rutgers New Jersey Medical School, Newark, NJ 07103, USA.
[3] KBR, Space Biosciences Division, NASA Ames Research Center, Moffett Field, CA 94035, USA.
[4] AI MATRIX Consortium, Department of Biomedical Engineering, University of Texas, San Antonio and UT Health Sciences, San Antonio, TX 78249, USA.
[5] Biological Systems and Engineering Division, Lawrence Berkeley National Lab, Berkeley, CA 94608, USA.
[6] DOE Agile BioFoundry, Emeryville, CA 94608, USA.
[7] Joint BioEnergy Institute, Emeryville, CA 94608, USA.
[8] Department of Physiology and Biophysics, Weill Cornell Medicine, New York, NY 10065, USA.
[9] Office of the Center Director, NASA Ames Research Center, Moffett Field, CA 94035, USA.
[10] Bay Area Environmental Research Institute, NASA Ames Research Center, Moffett Field, CA 94035, USA.
[11] Universities Space Research Association (USRA), Space Biosciences Division, NASA Ames Research Center, Moffett Field, CA 94035, USA.
[12] Department of Chemistry, University of North Florida, Jacksonville, FL 32224, USA.
[13] Center for Individualized Medicine, Department of Surgery, Department of Quantitative Health Sciences, Mayo Clinic, Rochester, MN 55905, USA.
[14] Department of Emergency Medicine, Center for Space Medicine, Baylor College of Medicine, Houston, TX 77030, USA.
[15] Department of Microbiology and Immunology, Department of Otolaryngology, Head and Neck Surgery, University of California San Francisco, San Francisco, CA 94143, USA.
[16] The Gilroy AstroBiology Research Group, The University of Wisconsin - Madison, Madison, WI 53706, USA.
[17] Weill Institute for Neurosciences, Department of Neurology, University of California San Francisco, San Francisco, CA 94158, USA.
[18] Data Science Analytics, Georgia Institute of Technology, Lima 15024, Peru.
[19] Hasso Plattner Institute for Digital Health at Mount Sinai, Department of Genetics and Genomic Sciences, Icahn School of Medicine at Mount Sinai, New York, NY 10065, USA.

[20] Center for Health AI, Department of Biochemistry and Molecular Genetics, University of Colorado School of Medicine, Anschutz Medical Campus, Aurora, CO 80045, USA.
[21] Center for Health AI, University of Colorado School of Medicine, Anschutz Medical Campus, Aurora, CO 80238, USA.
[22] Department of Neuroscience, University of Minnesota, Minneapolis, MN 55455, USA.
[23] Department of Computer Science, College of Science, San José State University, San Jose, CA 95192, USA.
[24] Biorelate, Manchester, M15 6SE, United Kingdom.
[25] UC Space Health, Department of Orthopaedic Surgery, University of California, San Francisco, San Francisco, CA 94143, USA.
[26] Faculty of Dental Medicine and Oral Health Sciences, McGill University, Montreal, Quebec, H4A 0A9, Canada.
[27] Faculty of Medicine, Dept of Surgery and Cancer, Imperial College London, London, SW7 2AZ, United Kingdom.
[28] SymbioSeq LLC, NASA Johnson Space Center, Ashburn, VA 20148, USA.
[29] Center for Data Driven Discovery, California Institute of Technology, Pasadena, CA 91125, USA.
[30] Waitt Advanced Biophotonics Center, Chan-Zuckerberg Imaging Scientist Fellow, Salk Institute for Biological Studies, La Jolla, CA 92037, USA.
[31] Human Research Program Cross Cutting Computational Modeling Project, NASA John H. Glenn Research Center, Cleveland, OH 44135, USA.
[32] Institute for Quantitative Health Science and Engineering, Department of Biochemistry and Molecular Biology, Michigan State University, East Lansing, MI 48824, USA.
[33] Department of Urology, Department of Radiology, Stanford University School of Medicine, Stanford, CA 94305, USA.
[34] Low Exploration Gravity Technology, NASA John H. Glenn Research Center, Cleveland, OH 44135, USA.
[35] Universities Space Research Association (USRA), Human Research Program Cross-cutting Computational Modeling Project, NASA John H. Glenn Research Center, Cleveland, OH 44135, USA.
[36] Mortality Research & Consulting, Inc., Houston, TX 77058, USA.
[37] Logyx, Space Biosciences Division, NASA Ames Research Center, Moffett Field, CA 94035, USA.
[38] Computer Science, Statistics, and Health Policy, Johns Hopkins University, Baltimore, MD 21218, USA.
[39] ML, AI and Healthcare Lab, Bayesian Health, New York, NY 21202, USA.
[40] Biotechnology and Planetary Protection Group, Jet Propulsion Laboratory, Pasadena, CA 91106, USA.
[41] SPHES, Medical Faculty, King's College London, London, WC2R 2LS, United Kingdom.
[42] Department of Genetics, Stanford School of Medicine, Stanford, CA 94305 USA.
[43] Department of Preventive Medicine and Community Health, UTMB, Galveston, TX 77551 USA.
[44] Human Health and Performance Directorate, NASA Johnson Space Center, Houston, TX 77058, USA.
[45] Department of Biology, California Institute of Technology, Pasadena, CA 91125, USA.
[46] ISS National Laboratory, Center for the Advancement of Science in Space, Melbourne, FL 32940, USA.
[47] UAB Center for Computational Biology and Data Science, University of Alabama, Birmingham, Birmingham, AL 35223, USA.
[48] Department of Biomedical Informatics, Harvard Medical School, Harvard Data Science, Broad Institute of MIT and Harvard, Harvard University, Boston, MA 02115, USA.
[49] Space Biosciences Division, NASA Ames Research Center, Moffett Field, CA 94035, USA.

[+]Corresponding Author Information:
Sylvain V. Costes <sylvain.v.costes@nasa.gov

## Abstract

Space biology research aims to understand fundamental effects of spaceflight on organisms, develop foundational knowledge to support deep space exploration, and ultimately bioengineer spacecraft and habitats to stabilize the ecosystem of plants, crops, microbes, animals, and humans for sustained multi-planetary life. To advance these aims, the field leverages experiments, platforms, data, and model organisms from both spaceborne and ground-analog studies. As research is extended beyond low Earth orbit, experiments and platforms must be maximally autonomous, light, agile, and intelligent to expedite knowledge discovery. Here we present a summary of recommendations from a workshop organized by the National Aeronautics and Space Administration on artificial intelligence, machine learning, and modeling applications which offer key solutions toward these space biology challenges. In the next decade, the synthesis of artificial intelligence into the field of space biology will deepen the biological understanding of spaceflight effects, facilitate predictive modeling and analytics, support maximally autonomous and reproducible experiments, and efficiently manage spaceborne data and metadata, all with the goal to enable life to thrive in deep space.

## Introduction

Space biology research focuses on answering fundamental mechanistic questions about how molecular, cellular, tissue, and whole organismal life responds to the space environment. Biological stressors of spaceflight include ionizing radiation, altered gravitational fields, accelerated day-night cycles, confined isolation, hostile-closed environments, distance-duration from Earth[1], planetary dust-regolith[2], and extreme temperatures/atmospheres[3,4]. Moreover, spaceflight stressors are likely compounded and amplified with increasing time in space and distance from Earth[1,5]. Understanding, predicting, and mitigating these changes at all levels of biology is increasingly important, given the deep space exploration goals of the National Aeronautics and Space Administration (NASA) toward cis-Lunar and Mars missions. Ultimately, the goal of space biology research is to extend beyond an understanding of how extraterrestrial conditions affect life, and develop foundational knowledge to support life (how plants, crops, seeds, microbes, and animals can thrive in deep space), leading to bioengineered solutions for sustained life on the Moon, Mars, and during deep space missions beyond low Earth orbit (LEO)[6].

Space biology research leverages spaceflown and ground-analog experiments using model organisms to understand space impacts on increasingly complex life. Experimental models include unicellular organisms (e.g., prokaryotic, eukaryotic, yeast, fungi), tissue-on-a-chip models[7], invertebrates (e.g., *Drosophila melanogaster*, *Caenorhabditis elegans*, tardigrades), simple model plants (e.g., *Arabidopsis thaliana*), vertebrates (e.g., mice, rats, fish), and crops and edible plants[1,7,8]. Model organism research is key to translational science, with the resulting evidence influencing the direction of human health research and driving the design of life support systems[9].

At the molecular and cellular levels, space biology experiments seek to characterize all possible spaceflight-induced changes in cell morphology[10], development and differentiation[11], protein regulation[12], epigenetic processes[13], and gene expression[14], among others[15]. Organ-level modeling systems like tissue-on-a-chip models are used to study shifts in cellular organization and communication[16–18]. The current understanding of biological responses to spaceflight incorporates experimental evidence from a variety of data types along hierarchical biological levels, from molecular to single-cell to whole-organism (**Figure 1**). At the cellular level, six fundamental responses to spaceflight have been well characterized, including increased oxidative stress, DNA damage, mitochondrial dysregulation, and telomere length, as well as epigenetic and microbiome changes[1]. These responses have been linked to a variety of physiological effects, including cardiovascular dysregulation, central nervous system impairments, bone loss and immune dysfunction[1].

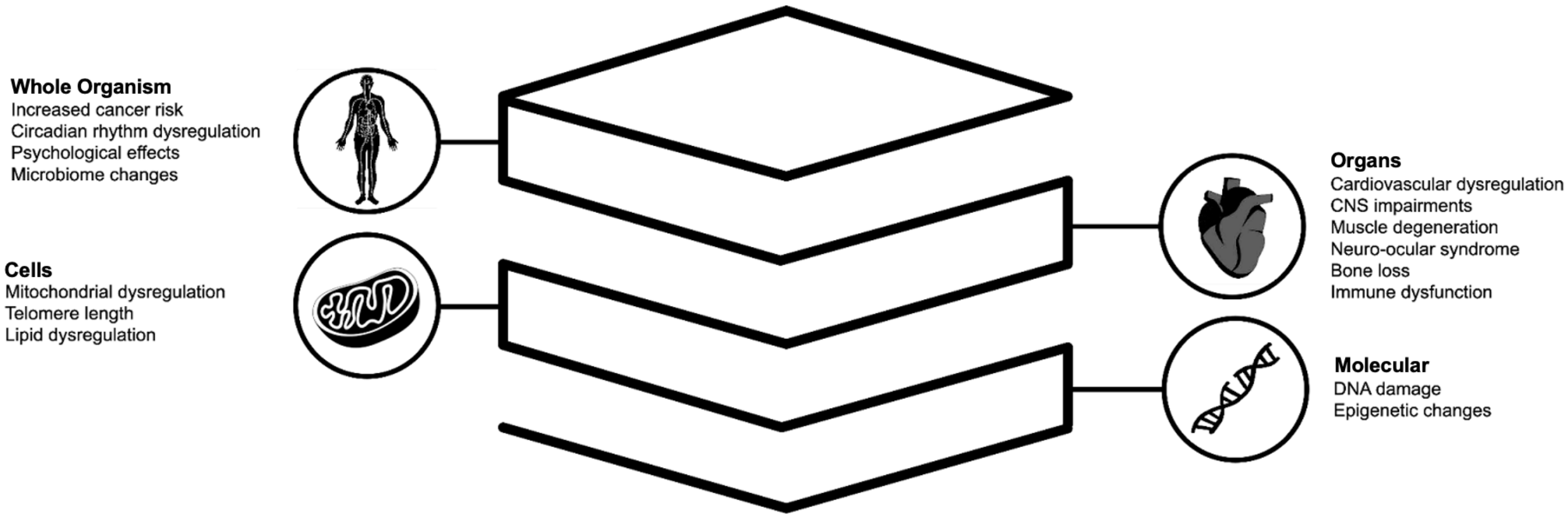

**Figure 1. Multi-hierarchical levels of space biological research and data.** Space biology research seeks to characterize the effects of spaceflight on living systems across hierarchical biological levels. Our current understanding of the biological responses to spaceflight incorporates multiple types of evidence at the cellular, tissue, and whole organism level.

The majority of current space biology knowledge originated through ground-analog experiments[19,20], from satellites in LEO[8], and from experiments on the Space Shuttle[21] or on the International Space Station (ISS)[22,23]. Despite decades of such experiments, essential biological knowledge gaps remain[4] for humanity to travel safely to the Moon, Mars, and beyond. LEO space experiments have most often been conducted by astronauts, which is inefficient and often necessitates small sample sizes. The design of more efficient and autonomous experimental research beyond LEO must not overburden the crew with tasks that could otherwise be automated, such as processing raw data. Furthermore, experiment hardware must be designed to be as low in mass and volume as possible, and to autonomously produce science results without guidance from crew or from Earth.

In this review, we present findings from the 'Workshop on Artificial Intelligence and Modeling for Space Biology' organized by NASA in 2021, which sought to map out the roles of artificial intelligence (AI), machine learning (ML), and biological-computational modeling in the field of space biology over the next decade. A separate article in this issue reviews workshop recommendations regarding these roles for astronaut health (Scott et al., 2021 [unpublished preprint]; Supplements 1 and 2).

Overall, the workshop concluded that fundamentally transformative approaches leveraging cutting edge data, coding, and AI-ML methods are needed for space biology, in addition to a general streamlining of biological experimentation and research activities in space to maximize validation of results (for both LEO experiments and beyond). These approaches must facilitate the generation and analysis of reproducible datasets which incorporate multiple types of measurements to achieve a comprehensive characterization of organismal responses to a variety of extraterrestrial conditions. Such datasets can then be used for robust predictive modeling of spaceflight responses at every biological level (Figure 1). We propose a widespread implementation of AI and ML methods at every level of space biology research (from ground to spaceborne research). This effort has the potential to revolutionize the breadth and depth of our knowledge base in two central ways: (1) enabling efficient, intelligent, and maximally autonomous experimental design and data collection in space research environments, where maximally implies as autonomous as possible while including an essential human link to assess, evaluate, and act on results and (2) by assisting management, analysis, modeling, and interpretation of current and future space biology datasets.

## Space Biology Experimentation

### Current Approaches in Space Biology Research

Space biology research has benefited from innovations in increasingly efficient and sensitive research technologies[24,25]. Next-generation sequencing platforms[26], big data frameworks and computational libraries for data storage, processing and analysis have led to the ability to conduct groundbreaking clinical studies with multiple types of 'omics data collected across thousands of samples[27,28]. Recent innovations in technologies such as single-cell sequencing, exosome sequencing, cell-free nucleic acid sequencing, spatial transcriptomics[29,30], and nanopore sequencing[31] have significantly broadened the potential for longitudinal characterization of cellular and genomic dynamics[29–35]. Findings derived from spaceflown experiments leveraging these technologies include retinal-ocular alterations[14,36], liver dysfunction[37,38], microRNA signatures[39], mitochondrial stress[40], gut microbiome alterations[41], and alternative splicing in space-grown plants[42].

However, it is difficult to leverage these technologies to their full potential in space, where workforce and resources are extremely limited. Most experimentation in space is expensive, time-consuming, and not automated, necessarily resulting in small experiments with few samples and replicates and significant variability due to differing sample handling procedures[43–45]. This makes ML analysis of space biological data difficult, as the models become under-determined and overfitted to the training data due to high dimensionality (tens of thousands of variables compared to tens or hundreds of data points), and technical batch effects make it challenging to combine datasets to gain higher sample numbers. To mitigate processing batch effects, NASA GeneLab provides open-source, uniformly processed multi-omics data from spaceflight and ground-analog studies, making space biology multi-omics data as AI-ready as possible[46] and providing a roadmap for the management of other spaceflight-relevant data (phenotypic, physiological, imaging, microscopy, and behavioral)[47]. Additional work is needed to establish widely-adopted standards for AI-readiness in these research domains[48,49]. These standards can then be leveraged by automated, on-board, end-to-end data collection and analysis platforms which we discuss in the following sections.

### AI-assisted Experiment Automation and Design

A comprehensive effort to streamline and automate biological experimentation in space is needed to generate the large-scale, high-quality, AI-ready, reproducible datasets required to meaningfully expand and validate our scientific understanding and knowledge base.

#### *Current Terrestrial Experimental Automation*

On Earth, basic molecular biology tasks such as pipetting, sequencing library preparation, cell culture maintenance, microscopy, quantitative phenotyping and behavioral change detection have already been automated in a variety of platforms[50–54]. Biofoundries apply high-throughput laboratory automation to generate thousands of strain constructs and DNA assemblies per week[55]. These advances now enable robust technical reproducibility across experiments, allowing researchers to isolate only the effects of relevant biological independent variables. However, these platforms still require significant personnel operation and hands-on time. Ideally, a fully automated experimental system for spaceborne research will integrate multiple robotic functions (e.g., pipetting + cell culture + microscopy photo capture and analysis + phenotyping + cell lysis and nucleic acid isolation + library preparation + sequencing + data analysis). The only human input required should be the initial setup of experimental parameters and the command to begin experimentation, and system-requested input when unexpected experimental outcomes are observed.

### *Current and Potential Spaceflight Experimental Automation*

Currently there is limited automation for biological data collection and analysis in spaceflight, although progress has been made, particularly in automating spaceborne biological image acquisition: for example, a real-time multi-fluorescence cell culture microscope was established on the ISS[56], and *Arabidopsis* response to microgravity was live-imaged by confocal microscopy[57]. A deep learning approach for automated cell segmentation based on massive crowdsourced annotation libraries was recently developed, which could be leveraged to greatly expedite *in situ* deep space knowledge discovery[58]. One possibility for AI and automation in space would be to move from the current, manual analysis of ISS rodent behavioral video[59], to an ML-based analysis of ambulatory, sensorimotor, and behavioral spaceflight effects[60,61]. Another possibility would be to leverage Natural Language Processing with Vision-Transformer models to develop platforms for automatic, real-time image descriptions and labeling[62,63].

Another area of expanding automated and AI data capture and analysis in spaceflight is ocular/retinal imaging. IDx-DR, an AI-enabled analysis platform for detection of diabetic retinopathy in retinal images, is one of the few FDA-approved AI-based methods[67]. This indicates potential feasibility of AI-based methods to detect space-related pathologies such as spaceflight-associated neuro-ocular syndrome (SANS), a high-priority ocular/visual risk to long-duration microgravity missions[68,69]. Currently, informative changes in vascular branching of the retina and other tissues can be mapped and quantified by NASA's AI-enhanced Vessel Generation Analysis (VESGEN) software[69]. Real-time detection of experimental results and pathologies in spaceflight could be enabled by full integration of VESGEN with computer-supported ophthalmic ocular coherence tomography (OCT) and OCT-angiography (OCT-A), which have recently been updated on the ISS for monitoring SANS[70], and are increasingly miniaturized[71] and AI-integrated[72]. Fundoscopy, OCT and OCT-A are now available for real-time, longitudinal imaging of small animals[73], which would greatly expand experimental capabilities[74].

Recent years have seen successful sequencing of nucleic acids aboard the ISS, facilitated by a long-read sequencer (Oxford Nanopore Technologies)[31,64–66]. Predictably, significant testing and adjustment was required for the sample loading and sequencing procedure due to the effects of microgravity on liquid dynamics[65], illustrating the investment required to automate complete experimental procedures in space, but providing a powerful example for transitioning state-of-the-art research capabilities to space.

### *Looking to the Future: Self-Driving Labs*

Automated experimentation in space will enable spaceflight-ready "self-driving labs"[75] that employ AI in a closed-loop system to produce new knowledge and optimize experimental design based on data collected in previous experiments (**Figure 2**). The AI system has the capability to choose the hypothesis to be tested and the parameters for the next experiment[76]. In the last decade, significant advances in several research areas have made self-driving labs possible[77–82]. We now have the ability to automate many biological processes using state-of-the-art microfluidics chips for optics, imaging and robotics[83–87].

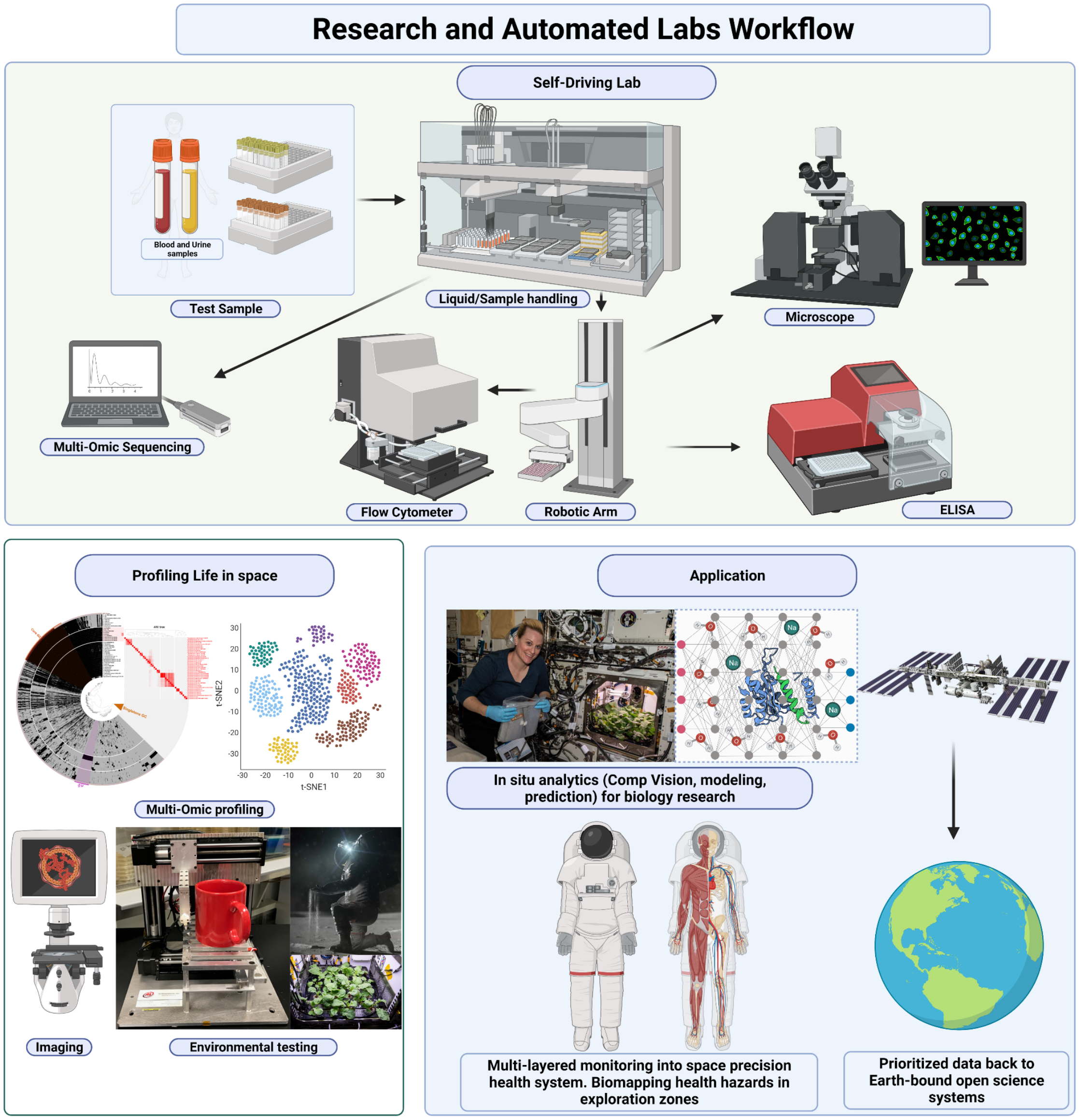

**Figure 2. Self-driving labs are automated experimental platforms with AI closed-loop control for knowledge gain and experimental design.** In spaceflown research programs, implementation of self-driving labs will aid comprehensive characterization of the effects of spaceflight on living systems, ultimately feeding research findings into applications such as *in situ* analytics, Earth-based open science research programs, and precision astronaut health systems.

The ultimate goal of developing and employing automated, AI-supported bioexperimentation systems such as self-driving laboratories should be to inform autonomous Precision Space Health systems that provide decision support for crew health management during LEO, cis-Lunar, and Mars missions (Scott et al 2021 [unpublished preprint]). As automated experimentation becomes more widely available, the space biology field should shift to conducting longitudinal studies[88], characterizing physiological changes over the duration of an entire mission. These longitudinal data will help identify biomarkers from various physiological, molecular and microbial systems that can be integrated to create individualized baseline models for humans and other organisms. Monitoring in-flight changes to these biomarker signals will help predict and prevent adverse organismal health outcomes, and predict how different organisms will react to spaceflight conditions.

## Data Frameworks and Algorithms

### Data Management for AI-Readiness

Raw biological data can be complex, sparse, and heterogeneous, and is therefore not typically ready for AI applications. Biological measurements relevant to a single scientific question may be discrete or continuous, qualitative or quantitative, single- or multi-dimensional, incomplete, highly descriptive (e.g. the appearance of cells in culture), and unstructured (particularly for phenotypic and behavioral data). Different experimental practices between facilities and researchers manifest as biases in the data, complicating integration of data from various experiments into a unified platform.

For space biology, the NASA GeneLab open-source data repository[46] has attempted to transform available 'omics data into AI-ready formats. However, significant manual work remains to process other types of data into AI-ready formats[47]. In order to best leverage all available data, the space biology field needs to invest in tools to perform automated conversion from existing, non-AI-ready formats into AI-ready formats. To facilitate this, the community needs a set of standardized ontologies and data formatting guidelines for space biology (for example, the inclusion of a datasheet to describe each dataset[89]). These standards can then inform experimental design to ensure that data from future missions are generated in an AI-ready format.

A key part of these data standards is the establishment of uniformly used vocabularies that are grounded in common conceptualizations (i.e. ontologies), which increase data discovery and reuse. Biomedical ontologies have existed for over 50 years and many are in widespread use[90,91], but no single ontology includes foundational concepts in the space biology domain (e.g., specially-developed experimentation hardware types, space environment types and parameters, etc). The space biology community should focus efforts on developing one or more such ontologies to standardize metadata with respect to space-relevant concepts and data structures (similar to the recent open-source, "Radiation Biology Ontology" publication by NASA GeneLab and STOREDB[92]).

Automated, AI-assisted data harmonization and dataset curation will be a critical part of advanced Space Biology research architectures like the one shown in **Figure 3**. Such architectures must be designed to support the entire experimental process: from investigation management, to experiment execution, to data publication, through Open Science[93] data repository submission, with appropriate security and governance measures to guarantee protection of private data resources. Investigators will be supported by an embedded digital experiment notebook to preserve experimental parameters and analyses, with link-out capability to approved, external data resources for seamless integration with research data. Use of space biology metadata ontologies can support automated harmonization across the wide spectrum of organisms studied, equipment used, and experimental designs. Because space biology datasets cover a range of modalities, each requiring distinct data processing, such advanced architectures must include a suite of metadata and data acquisition, and data processing tools. The proposed environment is similar to the successful virtual observatory paradigm[94,95] in the planetary sciences, and effective methodology transfer from planetary science to biomedical research has already occurred between NASA's Jet Propulsion Laboratory and the National Cancer Institute's Early Detection Research Network[96].

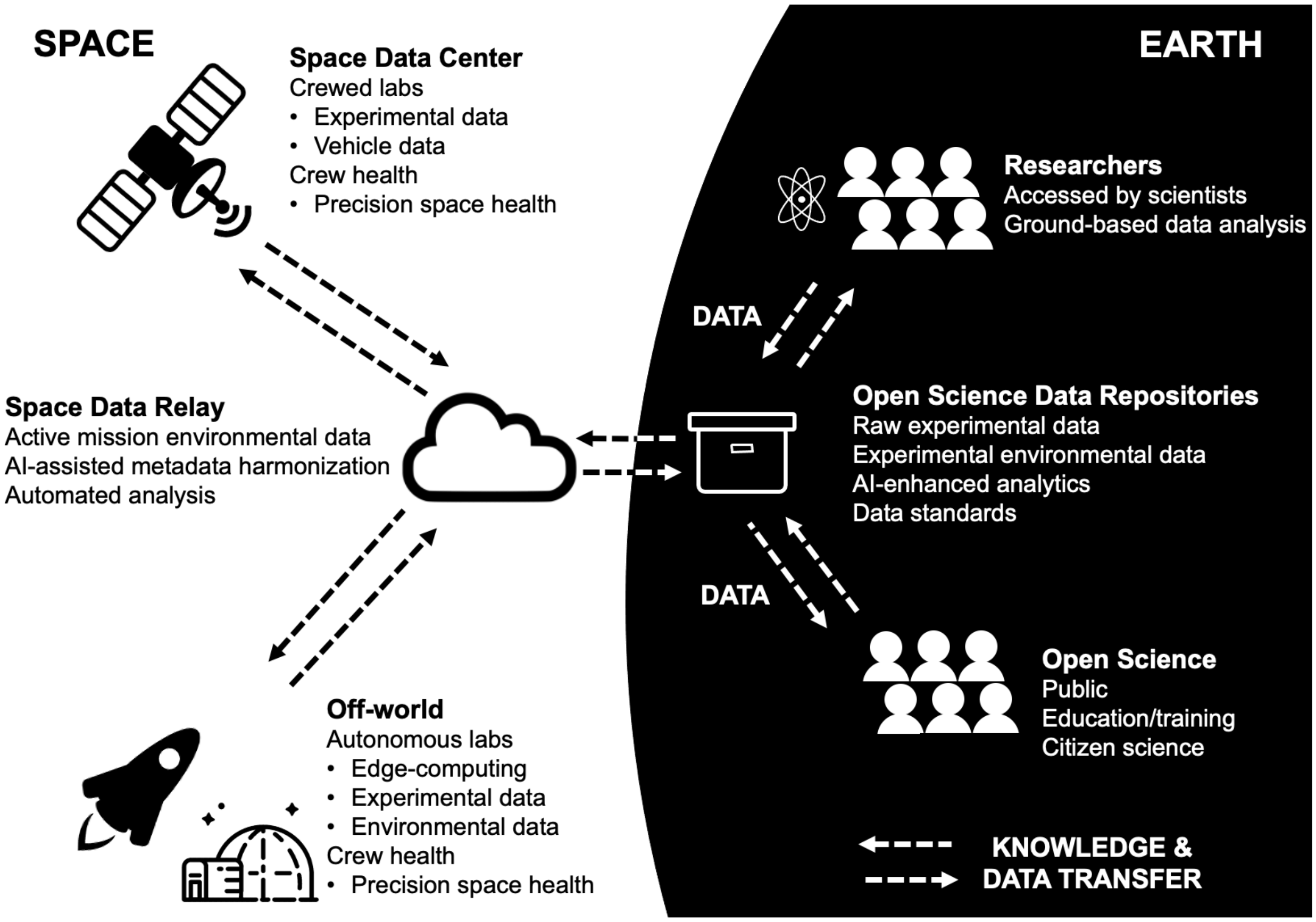

**Figure 3. Deep space biological and biomedical data collection and transfer.** The diagram shows the data and information flow in which a cloud-based data management environment serves as the nexus between space-based data and research and Earth-based researchers and analysts, enabling Open Science access to data and analytics and facilitating preparation of AI-ready datasets.

Full utilization of such an environment would ensure that all newly generated datasets are AI-ready, and facilitate conversion of previously generated datasets into AI-ready formats. Additionally, embedded Open Science capabilities will enable broad data sharing and reuse, and avoid metadata decay and long-term data maintenance issues. A similar data management tool was implemented recently by the National Institute of Standards and Technology, to address data harmonization and standards for their principal investigator and research community[97]. Ultimately, such an architecture will be cloud-based and linked to in-flight data acquisition systems, and eventually deep space communication for critical data downlinks.

Finally, it will be important to establish and adopt robust dataset readiness metrics in order to aid AI-modeling researchers in understanding the applicability of various datasets. Technology Readiness Levels have already been proposed for ML methods[98]; such metrics, if applied to datasets, could be useful for understanding the AI-readiness of space biological data. Moreover, a Bronze/Silver/Gold reproducibility standard has been proposed for life science ML workflows[99]. A similar standard could be implemented for ML analysis of space biological data in order to ensure reproducibility and confidence in results. These standards would be tailored for different ML methods.

## Models and Algorithms

Space biology combines the complexity of the biological and medical fields with an entirely new dimension: extended spaceflight in environments not known on, or very different from, Earth. Therefore, it is important that we consider the development of AI/ML algorithms specifically designed for data collected in novel space constraints and environments.

### *Organizing Data*

Standardized space biology ontologies will provide the opportunity to construct knowledge graphs[100,101] (KGs) compatible with space biology. These KGs will incorporate and model causal relationships using ontological content and space data, enabling the inference of physiological responses to experimental perturbations from multi-omic, phenotypic, imaging, and environmental-telemetry data. An existing relevant KG is the NSF-funded, UCSF-developed SPOKE (Scalable Precision Medicine Oriented Knowledge Engine)[102], which is linked to ~30 biomedical, chemical, molecular, and pharmaceutical databases[103]. Analysis of transcriptomic spaceflown mouse data using SPOKE identified spaceflight-induced physiological changes similar to terrestrial clinical conditions, consistent across multiple tissue types, demonstrating the utility of KG-based systems for furthering our understanding of space biology[102].

### *Interpreting Biological Data*

Explainable AI (xAI) provides a human-readable explanation of the evidence and rationale for predictions and recommendations, particularly important in biomedical research[104–108]. As the ultimate goal of space biological research is to establish predictive characterization of spaceflight effects on human astronaut health through translation from model organisms, all aspects of AI development in this field must fully embrace and incorporate xAI practices as well as post-hoc explainability and model interpretability with tools such as LIME[109] and SHAP[110].

### *Extrapolating Beyond Data*

Inferring the effects of space environments on human physiology from model organisms is nontrivial. Transfer learning, an ML method in which knowledge from models trained in one system is applied to a different system, has been used to map insights from mouse to human[111]. The space biology field should increase its application of transfer learning to leverage the huge amounts of spaceflown and radiation-exposed animal model data for insights into mammalian-human physiological responses to spaceflight. Further, transfer learning can be implemented in settings beyond LEO where models pretrained on larger, Earthbound datasets can be used to find insights in space biology research data generated during spaceflight.

### *Generating New Data*

We recommend the creation of a collection of generative models (model zoo) that have been pre-trained for each of the main types of space biology data; similar to the Kipoi repository for pre-trained models in genomics[112]. These models, which typically use generative adversarial (GANs)[113] or variational autoencoder architectures (VAEs)[114], can be used to produce synthetic data to validate the performance and translatability of new and existing space biology AI/ML modeling methods. These models should be engineered to support translation training for specific validation needs. For example, the EGRESS model generates symptomatic ECG signals that resemble the output of an astronaut wearable device, providing a large dataset of realistic synthetic data upon which to train models for astronaut health monitoring[115,116]; and GANs have been used to generate synthetic DNA[117] and RNA[118] sequencing data.

Generative models can also provide powerful solutions for data mapping: generating data based on source data which is often dimensionally smaller than the target. For example, recent research indicates that VAEs can translate ECG readings into an activation map that re-create the electrical activity of the heart; this can provide critical new insight into the health of specific heart regions[119]. The pretrained nature of these models is especially valuable for domains which are compute intensive, such as image processing[120,121], since the models in the zoo are designed to require relatively small amounts of additional training in order to adapt (or transfer) them to adjacent or subset problem spaces[122].

### *Scaling Computation*

An important finding in this workshop was the recognition of limited computational resources in space. Several solutions should be explored for different space-based applications. Edge computing brings computing resources to where the data are generated and could facilitate analysis in remote locations[123]. When analysis does need to occur on Earth, ML-based compression algorithms[124] could reduce the amount of data that need to be downlinked. Federated learning, which updates a global model while training smaller models on separate datasets[125], may be a promising strategy for processing automatically generated data from spacecraft and Lunar-Mars habitats. For example, CRISP (Causal Research and Inference Search Platform) is an ensemble ML method for biological causal inference[126], which now includes the capability for federated model training based on Intel's open-source OpenFL framework[127].

### *Next-Generation Models*

Our discussion so far has focused on both classical and contemporary AI methodologies. The next generation of AI models, neuromorphic computing, will closely mimic human neural structure and cognition in order to understand and predict outcomes in novel situations without context[128]. Neuromorphic processors are extremely small and light, making them ideal for space transport. This technology can be developed for deep space knowledge discovery, maximally autonomous platforms, and mission support. Space biological data is often collected in environments with limited analogous or comparable settings on Earth. The lack of context makes classical inference difficult, but incorporation of neuromorphic computing tools will allow us to learn novel insights from these datasets. Further, these next-generation models can be used to study and predict the function of extraterrestrial organisms (e.g., in studying extremophile physiology).

Beyond human neural computation, we should invest in development of biologically-informed deep learning algorithms[129]. In terrestrial biomedicine, such algorithms have been developed for biomedical capabilities such as patient stratification and drug response prediction[130–132]. Space biology could benefit from development of similar algorithms inspired by the biomorphic computing of known or newly discovered organisms in extreme environments. For example, immortal jellyfish are known to continuously regenerate their neural networks in anoxic or very low oxygen environments in the deep ocean[133]. Inspiration can be found in extreme biological neural compute methods to develop novel AI methods which may be more appropriate for learning about extreme environments like those found in space.

## AI-Informed Prediction and Countermeasure Design

The ultimate goal of space biological research is to be able to predict the effects of spaceflight at all physiological levels within diverse living systems, then develop the building blocks to support life, and bioengineer the foundations for sustained life beyond Earth. Such predictive modeling and bioengineering will only be possible once we are able to model all parts of living systems, introduce perturbations, and measure genetic, cellular and physiological outcomes longitudinally.

### Digital Twins

Building on automated, robotic and longitudinal data capture capabilities, space biology research will benefit from the development of predictive models of whole organisms ("digital twins"), which integrate multi-scale mechanistic mathematical modeling of an entire complex organism; from genes to cells to tissues to organs[134–136]. There now exist whole-cell computational models of microbes *Mycoplasma genitalium*[137] and *Escherichia coli*[138] for cellular predictions, and the ongoing Physiome Project develops mathematical models of the human body, from cells to tissues to organs, integrating chemical, metabolic, cellular and anatomical information[139,140]. Such models could integrate microbial-host cell interactions and environmental coupling data to predict responses to microbial population change or environmental perturbations. It will be important to identify an appropriate set of reference organisms (bacteria, eukaryote, archaea, viruses) as targets for high-fidelity digital twin models, which could be used to predict biological response under diverse extraterrestrial environments.

Digital twins will be critical for advancing several goals relevant to human space exploration. They will be useful for predicting the effects of spaceflight hazards on human physiology and for designing interventions or predicting the outcomes of countermeasures. They will enable the development of predictive models that can be personalized to individual human astronauts based on unique differences in genetics or physiology. Moreover, they can serve as useful scaffolds for extrapolation across organisms of different species through transfer learning.

### Cell and Tissue Engineering

Understanding the physiological effects of spaceflight stressors on human health will drive a need for biomedical interventions that can normalize or optimize astronaut physiology, including engineered cells or tissues (e.g., radiation-resistant skin, engineered retinas optimized for space-induced changes in air pressure, enhanced bone and muscle density, cells with increased DNA repair pathways, etc.).

AI will be required for the scalable design of engineered cells and tissues. Currently cell and tissue engineering activities are predicated on a design-build-test-learn (DBTL) cycle, which requires tremendous human input and time. AI is beginning to transform these bioengineering activities by automating and scaling DBTL cycles[82]. For example, ART (Automated Recommendation Tool)[141] is an AI-powered algorithm that leverages probabilistic modeling to recommend bioengineering interventions that can maximize production of useful metabolic engineering products. Methods like ART will soon enable the design of entire gene circuit architectures, given the growing availability of highly predictive computational models of mammalian gene circuits[142] and demonstrations that AI can bridge the gap between DNA sequences and engineered functions[143,144]. These approaches will be important for designing cells and tissues to perform functions unique to space biology.

Validation of the safety and efficacy of engineered cells and tissues will be a critical step toward deployment. AI will be needed to understand and predict the cell differentiation and cell fate dynamics of engineered tissues once they are implanted in human astronauts: for example, CellNet[145,146] is a network biology platform that was developed to evaluate stem cell differentiation states relevant to target cell types. This platform has now been modernized to accommodate RNA sequencing data and has been useful for assessing cell and tissue engineering methodologies[147]. Similarly, CellOracle[148] leverages mammalian gene regulatory networks to predict gene expression dynamics following genetic manipulations, and Enformer[149] leverages deep convolutional neural networks to predict gene expression from DNA sequences. The advent of single-cell RNA sequencing has driven several AI-powered algorithms for analyzing cellular dynamics and inferring cell fate trajectories[150]. New "logical switches" on customized T-cells can create entirely new kinds of immunotherapies, and these tools could eventually be used to target specific aberrant receptors in damaged cells[151]. Methods such as these will be important for benchmarking and evaluating engineered cells and tissues.

## Discussion

Our current understanding of the multi-tiered physiological effects of spaceflight stressors on a mammalian organism is derived from a tiny amount of human astronaut data and hundreds of small, expensive, model organism biological experiments performed manually during a variety of spaceflight missions. Workshop participants agreed that to advance space biology research as a field, a paradigm shift is necessary from the current manual, single-experiment paradigm, into a new era of biological research conducted in space facilitated by robotic automation, AI-driven experimental design and analysis. Workshop participants envisioned an AI/ML space biology research lifecycle, with a data management environment and appropriate AI/ML methods facilitating the acceleration of research findings and ultimately powering widespread flight data acquisition and precision health support for astronaut health (**Figure 4**).

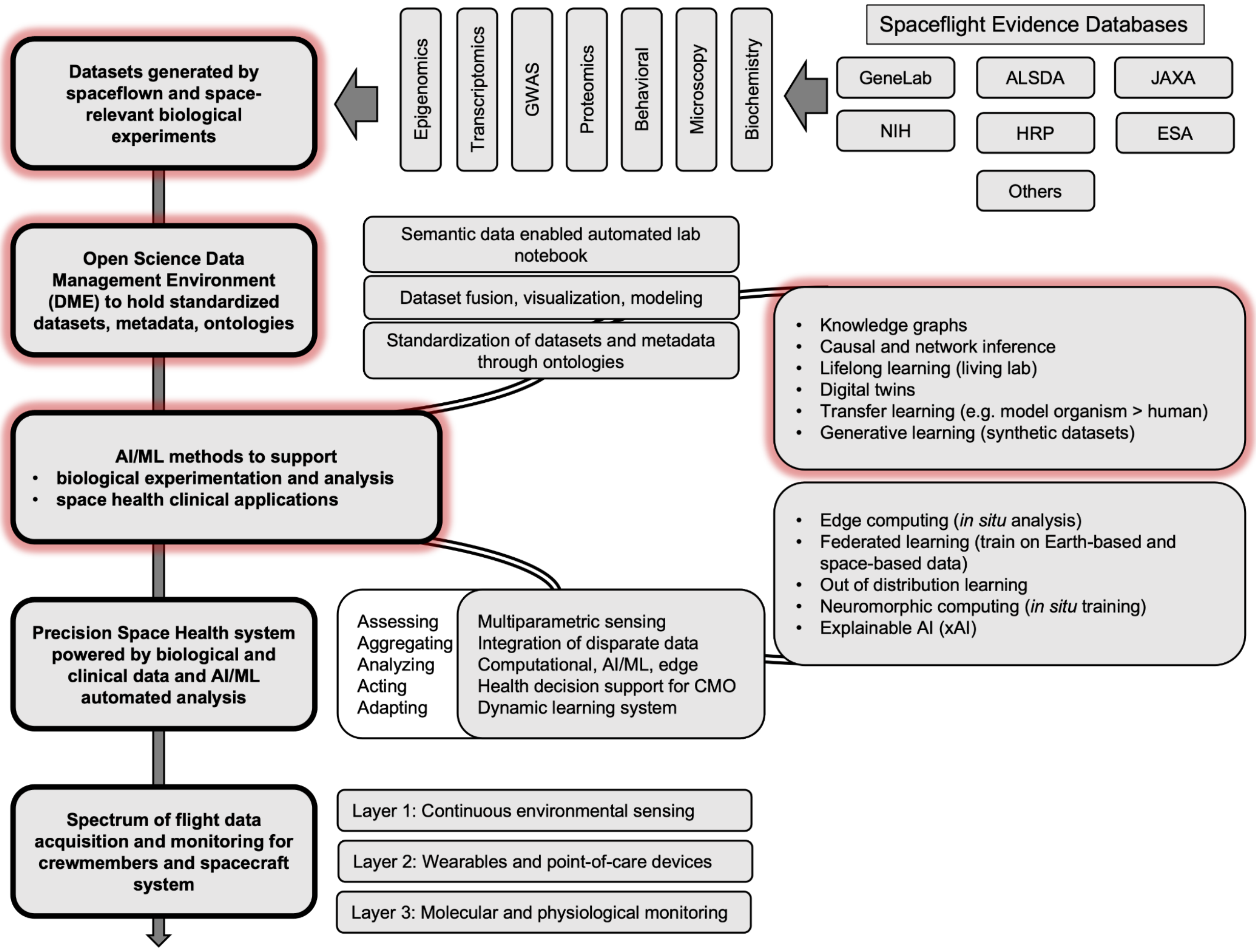

**Figure 4. Space biology and precision space health AI/ML life cycle.** The lifecycle of AI/ML in space biology research enables full utilization of Open and FAIR space biological datasets through a Data Management Environment and space biology-appropriate AI/ML models, to ultimately facilitate a Precision Space Health astronaut support system.

## Make AI Space-Ready

While much of the automation discussed above is already in use terrestrially, it is important to note that these hardware and software are not immediately suitable for spaceflight research, and steps must be taken to convert and develop these automated systems for use in-flight, likely following a logical progression of deployment(sub-orbital, ISS/LEO, Lunar Gateway, Lunar surface, and Mars transit, Martian surface, etc.). These spaceflight-ready automated systems will enable cost-effective collection of large biological data in difficult or constrained conditions. Moreover, workshop participants agreed that the next step is to couple automation with AI-assisted or AI-driven hypothesis generation and experimental design to facilitate the automatic generation of biological insights over time without the need for human input and expertise (self-driving labs). By benefiting from the high reproducibility of machines, we envision a future where automatic data and metadata acquisition will be complete and unambiguous such that AI methods will be able to accumulate such information, constantly learning from them and making the whole greater than the sum of the parts.

## Data Standards and Fields of Research

The future of space biology experimentation envisioned during this workshop will only be possible through widespread adoption and implementation of standards for generating and maintaining data and metadata from automated and AI-driven systems. It will be vitally important to develop a set of guidelines for generating AI-ready, machine-readable data from every space biology experiment, to facilitate open access AI/ML-assisted data analysis and reuse. This must include concerted efforts to develop and maintain space biology vocabularies and ontologies that can be leveraged for automated reasoning, as well as top-down motivation to adopt standardized data management facilities.

## Adapting and Developing AI Methods for Space Biology

Workshop participants discussed existing AI/ML methods, models and algorithms, and agreed that the next decade of investment must include a focus on adaptation and implementation of existing methods with a specific focus on space biology. Approaches such as transfer learning and generative modeling hold great promise for space biology research, but care must be taken to adapt these methods with the constraints of spaceflight in mind. Further, novel AI/ML approaches with high potential for space biology were discussed, including neuromorphic computing. Due to limited bandwidth in space, our efforts should focus on developing multi-faceted solutions including pre-training lightweight models on larger Earthbound datasets, federated training, edge computing, and on-board processing.

In the next decade, investment in AI/ML research design and analysis promises to revolutionize the way biological research is performed in space. Integration of automated and self-training systems will enable the hands-off and reproducible generation of huge cutting-edge imaging and multi-omics datasets, ready to be mined by next-generation AI/ML space biology tools. Workshop participants agreed that the key to this future involves the creation of multi-disciplinary teams with statisticians, biologists, modeling experts and hardware developers. Such partnerships will facilitate the experimentation and data analysis necessary to fully understand and begin to predict and mitigate the stressors of spaceflight, and enable life to thrive in deep space.

# Recommendations and Conclusion

Workshop participants agreed on several key development areas required for AI/ML technologies to be maximally integrated into space biological research:

- AI-assisted automated experimental platforms and self-driving labs in spaceflight
- Standards for AI-readiness for diverse biological data types
- Open Science data management environment to standardize, organize, and analyze all space biological data, both Earth- and space-generated
- Adaptation of existing AI/ML methods and development of new methods best-suited for space biology data
- Digital twins for prediction of biological systems response to space environments

Full adoption of AI/ML methods in space biological research is an endeavor that will span the next decade, but will ultimately revolutionize the way we perform experiments and analyze data for knowledge gain. The developments discussed in this paper will finally enable us to gather the necessary data and tools to build a comprehensive characterization of the biological responses of living systems to myriad diverse space environments. This knowledge base will be essential to facilitate NASA's goals of Lunar, Martian and deep space missions, as we will be able to predict and mitigate adverse effects at all biological levels.

## Acknowledgments

We thank all June 2021 participants and speakers at the “NASA Workshop on Artificial Intelligence & Modeling for Space Biology.” Thanks to the NASA Space Biology Program, part of the NASA Biological and Physical Sciences Division within the NASA Science Mission Directorate; as well as the NASA Human Research Program (HRP). Also, thanks to the Space Biosciences Division and Space Biology at Ames Research Center (ARC), especially Diana Ly, Rob Vik, and Parag Vaishampayan. Thanks also to the support provided by NASA GeneLab, and the NASA Ames Life Sciences Data Archive. Additional thanks to Sharmila Bhattacharya, NASA Space Biology Program Scientist; Kara Martin, ARC Lead of Exploration Medical Capability (an Element of HRP); as well as Laura Lewis, ARC NASA HRP Lead.

Funding: S.V.C. is funded by NASA Human Research Program grants NNJ16HP24I. S.E.B holds the Heidrich Family and Friends endowed Chair in Neurology at UCSF. S.E.B. also holds the Distinguished Professorship I in Neurology at UCSF. S.E.B is funded by an NSF Convergence Accelerator award (2033569) and NIH/NCATS Translator award (1OT2TR003450). G.I.M was supported by the Translational Research Institute for Space Health, through NASA NNX16AO69A (Project Number T0412). E.L.A. was supported by the Translational Research Institute for Space Health, through NASA NNX16AO69A J.Y. is funded by NIH grant # R00 GM118907. C.E.M. thanks NASA grants NNX14AH50G and NNX17AB26G. This work was also part of the DOE Agile BioFoundry (http://agilebiofoundry.org), supported by the U.S. Department of Energy, Energy Efficiency and Renewable Energy, Bioenergy Technologies Office, and the DOE Joint BioEnergy Institute (http://www.jbei.org), supported by the Office of Science, Office of Biological and Environmental Research, through contract DE-AC02- 05CH11231 between Lawrence Berkeley National Laboratory and the U.S. Department of Energy. SVK is funded by the Canadian Space Agency (19HLSRM04) and Natural Sciences and Engineering Research Council (NSERC, RGPIN-288253).

## Authorship Contributions

All authors contributed ideas and discussion during the joint workshop writing session or were speakers at the: “NASA Workshop on Artificial Intelligence & Modeling for Space Biology.” L.M.S., R.T.S., and S.V.C. prepared the manuscript. All authors provided feedback on the manuscript.

## Supplement 1 - Workshop Overview

To explore the future role of AI-modeling in space biology and health, NASA held a workshop in June 2021. The workshop was organized by the NASA Space Biology Program within the Biological and Physical Science Division, part of the NASA Science Mission Directorate. The NASA Human Research Program also supported and participated. The workshop gathered a cohort of external-to-NASA AI-modeling subject matter experts (SME) in the fields of digital health, computer science, bioinformatics, medicine, microbiology, biomedical imaging and computational biology.

The workshop's first day was organized to educate the AI-modeling SME cohort regarding: (1) long-term required biological and health capabilities needed for Lunar, Martian and deep space missions, (2) statuses of relevant data repositories, their content-structure and overall workflow of the current data resources to be mined-utilized, (3) current space-relevant biological AI, modeling and data science projects, (4) the unique statistical, data volume, cross-comparison and logistical challenges of data pertaining to astronaut health and space biological sciences. Select 'central domain topics' guided the workshop:

- AI and Modeling for Knowledge Discovery: 'Omics and other Space Biological Data
- AI Applications in Imaging Space Biology Research Data (including Behavioral Analysis AI Tools for Space Data)
- Precision Medicine Utilization of AI
- Data Collection through Wearables, Sensors, Monitoring Hardware Systems and Integration with AI and Modeling Power
- Space Health Risk Predictions through AI, Modeling, Network Analyses
- Spaceflight Countermeasure Predictions Utilizing AI, Modeling, and Network Analyses
- AI Applications for Microbiology and Synthetic Biology
- AI Techniques and Translational Science Across Model Organisms and Species Toward Human Health

On the second day of the workshop, the SME cohort and space-related researchers outlined AI and modeling recommendations and concepts for the next decade in space biology and space health.

## Supplement 2 - Workshop Flyer

National Aeronautics and Space Administration

### Workshop on Artificial Intelligence & Modeling for Space Biology

Supporting the Next Phase of Space Exploration and Discovery

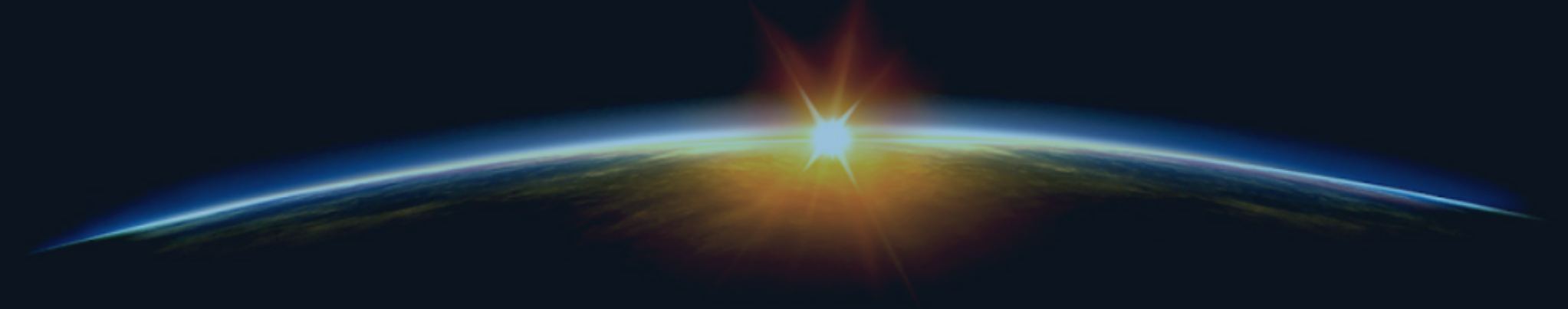

**What:** Please join us for this two-day workshop which will involve the participation of researchers and leaders in AI, modeling, biology, and medicine. Expertise is sought to outline a vision on where AI and modeling ought to be developed in the next decade towards enhancing space biological knowledge discovery and supporting health for space exploration missions.

**When:** Thursday-Friday, June 24-25th 2021

**Where:** Virtual Format

**Who is Organizing:** Hosted by NASA Biological and Physical Sciences (BPS) Division of NASA Science Mission Directorate, Biosciences Division at Ames Research Center

nasa.gov

National Aeronautics and Space Administration

**Purpose of Workshop:**
This NASA workshop will involve the participation of researchers and leaders in AI, modeling, biology, and medicine. Their expertise will be sought to outline a vision on where AI and modeling ought to be developed in the next decade towards enhancing space biological knowledge discovery and supporting health for space exploration missions.

Workshop will produce a White Paper for the NASEM-CBPSS and their Decadal Survey 2023-32. It also will produce a Workshop Summary & Recommendations in a peer-reviewed Journal.

Invited AI experts will help bridge the gap between technical space biology and health needs in the next decade, and the current status of our field, with all its various space data constraints, and science knowledge gaps.

Of note, this workshop will build upon the lessons, insights, and networks enabled by the May 2021 workshop on Artificial Intelligence hosted by the NASA Science Mission Directorate (SMD). The May 2021 SMD AI workshop covers all of NASA Science (Heliophysics, Earth Science, Biological, Physical, Astrophysics, etc).

This June 2021 Artificial Intelligence & Modeling for Space Biology workshop will focus on capturing the best practices and vision of AI capabilities specifically towards biological and life-science related health goals.

**Who is Organizing:**
Hosted by NASA Biological and Physical Sciences (BPS) Division of NASA Science Mission Directorate, Biosciences Division at Ames Research Center

**Workshop Organizing Team:**
Sylvain Costes, Kara Martin, Aenor Sawyer, Jenna Lang, Ryan Scott, Lauren Sanders, Dan Berrios, Amina Qutub

**Date:** Thursday-Friday, June 24-25th 2021

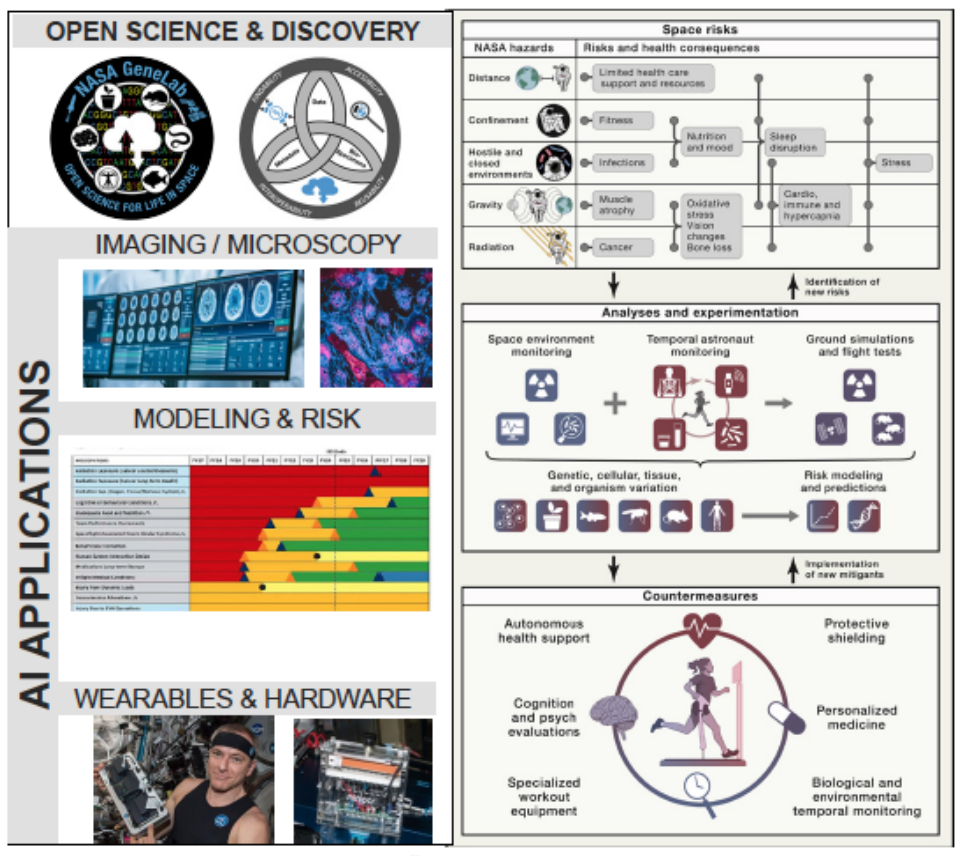

*Right, the space environment introduces several different hazards (distance, confinement, hostile/closed environments, radiation, and microgravity), which have numerous biological consequences and human health risks. Left, AI has been identified as a key field to be utilized towards accelerating biological knowledge discovery, imaging/microscopy, modeling/risk, and integration with various wearables and hardware systems.*

*(CREDIT: Right from Cell Press, https://doi.org/10.1016/j.cell.2020.10.050; Left top, Cell Press, https://doi.org/10.1016/j.celrep.2020.108441; Middle left, NASA Human Research Roadmap; Bottom left, CSA/NASA image with Astroskin and David Saint-Jacques; MiniPCR Machine, NASA from ISS)*

National Aeronautics and Space Administration

## Goals, Products, Format, Participants, Domain Topics

**Goals of the Workshop:**

- Identify AI technological needs for the next 10 years to address scientific gaps of space biology and health
- Develop strategic framework to fulfill gaps, emphasizing research consortiums, and funding mechanisms involving public and private partnerships
- Generate roadmap for AI and Modeling in space biology and health with key necessary capabilities and potential NASA campaigns

**Products:**

- White paper for NASEM-CBPSS and their Decadal Survey 2023-32, which workshop participants are encouraged to coauthor/cosign
- Pre-arranged peer-reviewed Journal Publication Summary

**Format:** Virtual workshop

**Length of workshop:** 2 days

- Day 1 – Overview, Keynotes, Learn from AI SMEs, State of Space Biology and Health Breakout Session 1, Writing
- Day 2 – Incubation Meeting of invited AI SMEs, Report from SME's Incubation Meeting, Report from Breakout Session 1, Breakout Session 2, Report from Breakout Session 2, Writing

**Participation:** 90 invitees. ~70 external AI-Data-Biology SMEs (both academic & industry). ~30 NASA internal (HQ, JSC, ARC, MSFC, GRC, etc.). All those interested in attending and contributing to the workshop are welcome.

## Central Domain Topics in our Workshop:

- AI and Modeling for Knowledge Discovery: 'Omics and other Space Biological Data
- AI Applications in Imaging Space Biology Research Data (including Behavioral Analysis AI Tools for Space Data)
- Precision Medicine Utilization of AI
- Data Collection through Wearables, Sensors, Monitoring Hardware Systems and Integration with AI and Modeling Power
- Space Health Risk Predictions through AI, Modeling, Network Analyses
- Spaceflight Countermeasure Predictions Utilizing AI, Modeling, and Network Analyses
- AI Applications for Microbiology and Synthetic Biology
- AI Techniques and Translational Science Across Model Organisms and Species Towards Human Health

## Vision of AI and Modeling in Our Field, Suggested References

### The Vision of AI and Modeling in Our Field of Space Biology:

The era of artificial intelligence (AI) has opened new possibilities for the fields of biological sciences, physical sciences, and medicine. Because AI techniques can automate scientific and analytical processes while reducing the burden of required time and resource constraints, there are high expectations for AI-enabled knowledge discoveries. In the context of space biology and the interests of NASA, the steady growth of quantitative biological data such as DNA sequencing and omics (genomics, epigenetics, proteomics, ...) is enabling AI applications for the prediction of disease risks, the identification of potential therapeutics, and the improvement of astronaut health in general. Also, the fusion of AI tools with the field of traditional biological modeling holds promise in quickening the pace of knowledge discovery. In addition, space biology ought to leverage the \ well-established usage of AI for image processing for knowledge extraction in diagnostic radiology, histopathology, immunohistochemistry, and cognitive-behavioral analysis from animal-based videos. AI will play a crucial role in the development of semi-autonomous health support for flight medical officers through health monitoring and integration with wearable sensors. In the near-future, assistance in medical diagnostics will be enabled by integrating state-of-the-art biotechnologies such as in situ genetic sequencing capability with health monitoring, for real-time personalized health-risk management. Lastly of note, by pushing AI technologies to their limit to address space biological and health-related challenges, myriad applications from this endeavor will be applicable on Earth and benefit terrestrial health substantially.

### Suggested Background References to Understand the Field:

- Cell Package: The Biology of Spaceflight
- Fundamental Biological Features of Spaceflight: Advancing the Field to Enable Deep-Space Exploration
- Evidence Reports from the NASA Human Research Program
- Research Methods for the Next 60 Years of Space Exploration
- FAIRness and Usability for Open-access Omics Data Systems
- NASA GeneLab: interfaces for the exploration of space omics data
- Policy Considerations for Precision Medicine in Human Spaceflight
- FDL, 2020, CRISP, Astronaut Health Technical Memo
- Integrating Spaceflight Human System Risk Research
- Knowledge Network Embedding of Transcriptomic Data from Spaceflown Mice Uncovers Signs and Symptoms Associated with Terrestrial Diseases
- From the bench to exploration medicine: NASA life sciences translational research for human exploration and habitation Missions
- Comprehensive Multi-omics Analysis Reveals Mitochondrial Stress as a Central Biological Hub for Spaceflight Impact
- FDL, 2019, Harnessing AI to support medical care in space, Technical Memo
- Considerations for Wearable Sensors to Monitor Physical Performance During Spaceflight Intravehicular Activities
- AI's role in deep space

National Aeronautics and Space Administration

## Confirmed Speakers

**Dr. Sylvain Costes**
Space Biosciences
Research Branch Chief (Acting)
GeneLab Project Manager
Lead of the Radiation Biophysics Laboratory
Senior Research Scientist, Code SCR
NASA Ames Research Center

**Dr. Lauren Sanders**
GeneLab Staff Scientist
NASA Ames Research Center

**Dr. Manil Maskey**
Senior Research Scientist,
Earth Science Data Systems
NASA Science Mission Directorate
NASA Headquarters

**Dr. Michael Snyder**
Stanford B. Ascherman
Professor and Chair,
Department of Genetics
Director, Stanford Center for
Genomics and
Personalized Medicine
School of Medicine,
Stanford University

**Dr. Kathleen Rubins**
NASA Astronaut and
Molecular Biologist
NASA Johnson Space Center

**Dr. David Van Valen**
Assistant Professor of Biology
and Biological Engineering
Caltech

**Dr. Greg Corrado**
Distinguished Scientist,
Google AI

**Dr. Amina Qutub**
Associate Professor
Department of Biomedical Engineering
University of Texas, San Antonio
Director, UTSA-UT Health Graduate Group
in Biomedical Engineering
Research Lead, AI MATRIX Consortium

**Dr. Casey Greene**
Professor, Department of
Biochemistry and Molecular Genetics
Director of the Center for Health AI
University of Colorado
School of Medicine

**Dr. Sharmila Bhattacharya**
Program Scientist for NASA Space Biology
Biological and Physical Science Division
NASA Headquarters

**Dr. Sergio Baranzini**
Department of Neurology,
Weill Institute for Neurosciences
University of California
San Francisco

**Dr. Jonathan Galazka**
GeneLab Project Scientist
NASA Ames Research Center

**Dr. Adrienne Hoarfrost**
NASA Postdoctoral Fellow
Rutgers University

**Dr. Patricia Parsons-Wingerter**
Biomedical Research Engineer
Low Gravity Exploration Technology (LTX)
NASA Glenn Research Center

**Dr. Mary Van Baalen**
Acting Chair, NASA Human Systems Risk Board
Space Medicine Operations Division,
NASA Johnson Space Center

**Dr. Robert Reynolds**
Baylor College of Medicine
Translational Research Institute for
Space Health

**Dr. Jessica Keune**
Deputy Lead of NASA's Life Sciences
Data Archive & Lifetime Surveillance
of Astronaut Health
Space Medicine Operations Division,
NASA Johnson Space Center

**Dr. Aenor J Sawyer, MD, MS**
Director, UC Space Health
Director, UCSF Skeletal Health Service
Department Orthopaedic Surgery, UCSF
Co-Director, UCSF Center for
Advanced 3D + Technologies

**Graham Mackintosh**
AI Projects Consultant
Bay Area Environmental Research Institute
NASA Advanced Supercomputer Division
NASA Ames Research Center, Code TN

National Aeronautics and Space Administration

## Agenda Day 1: Thursday, June 24 *all times PDT*

| | Session | Presenter / Panelists |
|---|---|---|
| 7:30 – 7:50 am | **Welcome**<br>Moving Space Biosciences to Knowledge-Based Platforms<br>Overview from 2021 Space Mission Directorate AI Workshop | **Sylvain Costes** NASA, BPS<br>**Lauren Sanders** NASA, BPS<br>**Manil Maskey** NASA SMD |
| 7:50 – 8:40 am | **Keynote**<br>AI Applications in Biosciences | **Mike Snyder** Stanford |
| 8:40 – 9:00 am | **Break** | |
| 9:00 – 10:00 am | **Keynote**<br>AI Applications in Space Biology | **Kathleen Rubins** NASA Astronaut |
| | **Learn From The Expert Mini-Symposium**<br>4 concurrent sessions | |
| 10:00 – 10:50 am | Single-cell biology in a Software 2.0 World | **David Van Valen** Caltech |
| | Frontiers of Machine Learning | **Greg Corrado** Google AI |
| | Artificial Intelligence, Systems Biology, Brain, and Health | **Amina Qutub** UT San Antonio |
| | AI in biology is powered by open data | **Casey Greene** University of Colorado |
| 10:50 – 11:10 am | **Break** | |
| | **Current State of Space Biology and Health – Resources and Challenges**<br>2 concurrent sessions | |
| 11:10 am – 1:00 pm | Space Biology and Physical Sciences Research | **Sharmila Bhattacharya** NASA Space Biology<br>**Sergio Baranzini** UCSF<br>**Jon Galazka** NASA, BPS<br>**Adrienne Hoarfrost** Rutgers University<br>**Patricia Parsons-Wingarter** NASA |
| | Human Health and Space Medicine | **Mary Van Baalen** NASA<br>**Rob Reynolds** Baylor<br>**Jessica Keune** NASA<br>**Aenor Sawyer** UCSF |
| 1:00 – 1:30 pm | **Break** | |
| 1:30 – 2:30 pm | **Breakout Session 1**<br>**Key Concept**<br>Knowledge and technology gaps in Central Domain Topics | |
| 2:30 – 2:50 pm | **Break** | |
| 2:50 – 4:00 pm<br>(Specific cohort of participants) | **Breakout Session Report Writing** | Organizers, moderators, and notetakers |

## Agenda Day 2: Friday, June 25 *all times PDT*

| | Session | Presenter / Panelists |
|---|---|---|
| 7:30 – 9:00 am<br>(Specific cohort of participants) | **Incubation Meeting for Specific AI-Data Science SME Attendees**<br>Identification of key needs and knowledge gaps<br>Development of vision for Decadal Survey white paper | Moderators<br>**Amina Qutub** UT San Antonio<br>**Lauren Sanders** NASA, BPS |
| 9:00 – 9:20 am | **Break** | |
| 9:20 – 10:40 am | **Incubation Panel Report and Breakout Report** | **Amina Qutub** UT San Antonio<br>**Lauren Sanders** NASA, BPS |
| 10:40 – 11:00 am | **Break** | |
| 11:00 – 11:10 am | **Breakout Session Kickoff** | **Graham Mackintosh** NASA |
| 11:10 am – 12:40 pm | **Breakout Session 2**<br>**Key Concept**<br>AI applications needed by space biology and health | |
| 12:40 – 1:10 pm | **Break** | |
| 1:10 – 1:40 pm | **Open Discussion** | **Sylvain Costes** NASA, BPS |
| 1:40 – 1:50 pm | **Instructions for Writing** | |
| 1:50 – 4:00 pm | **Breakout Writing Rooms (Optional)**<br>Central Domain Topic focus<br>Collation of content for Decadal Survey white paper and journal article | |

National Aeronautics and Space Administration

## Participants identify areas of expertise relating to “4 Quadrants” for breakout discussions

**4 Quadrants for All Participants:**
To organize discussions around AI, Modeling, space biology, medicine, the workshop organizers have developed a “4 Quadrants” framework to describe participant expertise. The areas of expertise within the quadrants are not meant to be all-inclusive, but rather to provide an overview of each quadrant. The organizers recognize that there are many ways to stratify the AI, modeling, biology, and medicine communities. The “4 Quadrants” framework was designed to focus discussion, disperse expertise across domains, and maximize the precious time attendees give to participation in the workshop.

**Purpose:**
The “4 Quadrants” framework will aid in distributing expertise evenly throughout the breakout discussions. These breakouts will have critical input to the white paper and journal paper deliverables.

**Actions:**
All participants will be asked to identify one or more main areas of expertise during registration, which will be used to stratify participants into quadrants.
Before the workshop, participants are asked to give some thought to examples of successes, challenges, and how to optimize the inherently inter-disciplinary and diverse teams required to leverage AI and Modeling towards space biology discovery and health.

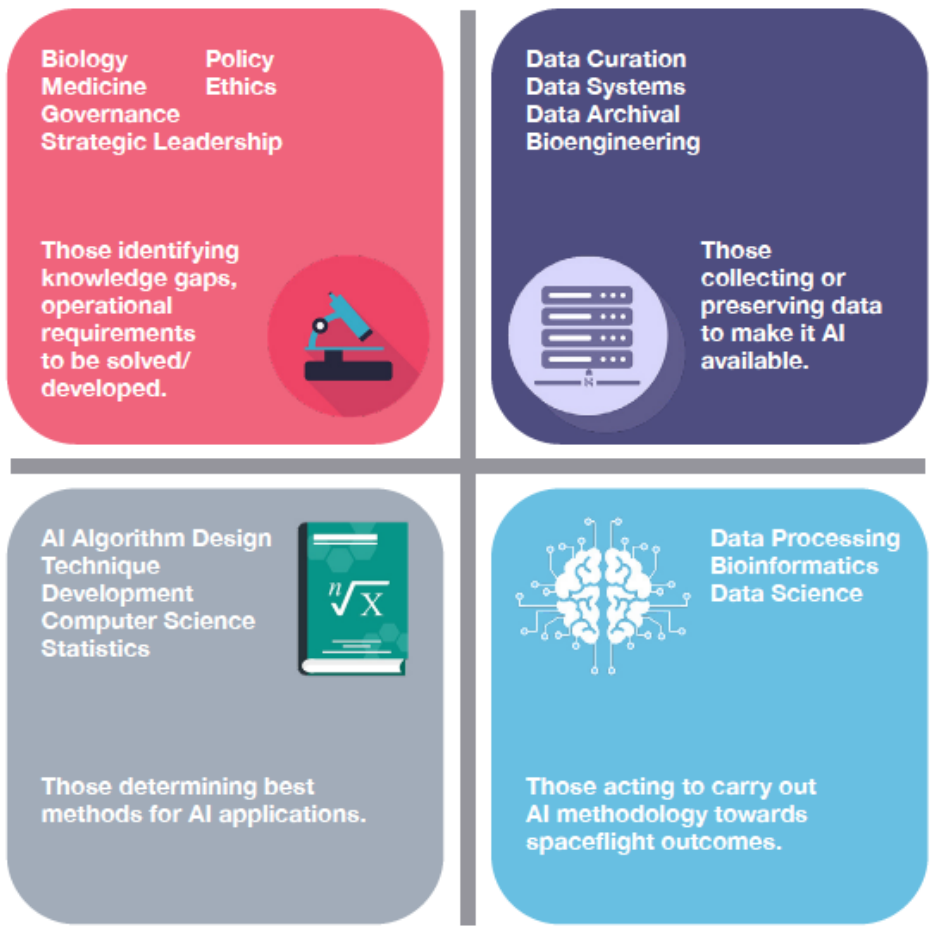